\documentclass[preprint,prd,nofootinbib,superscriptaddress,tightenlines,amsmath,amssymb]{revtex4}

\usepackage{bm}
\usepackage{color}
\usepackage{graphicx}

\begin{document}
\baselineskip=15pt \parskip=5pt

\vspace*{3em}

\preprint{KIAS-P16031}

\title{Flavor-Changing Higgs Decays in Grand Unification\\with Minimal Flavor Violation}

\author{Seungwon Baek}
\email{swbaek@kias.re.kr}
\affiliation{School of Physics, Korea Institute for Advanced Study, 85 Hoegiro Dongdaemun-gu,
Seoul 02455, Korea\smallskip}

\author{Jusak Tandean}
\email{jtandean@yahoo.com}
\affiliation{Physics Division, National Center for Theoretical Sciences, Hsinchu 300, Taiwan}
\affiliation{Department of Physics and Center for Theoretical Sciences,
National Taiwan University, Taipei 106, Taiwan\bigskip}


\begin{abstract}
We consider the flavor-changing decays of the Higgs boson in a grand unified theory framework
which is based on the SU(5) gauge group and implements the principle of minimal flavor violation.
This allows us to explore the possibility of connecting the tentative hint of the Higgs decay
\,$h\to\mu\tau$\, recently reported in the CMS experiment to potential new physics in
the quark sector.
We look at different simple scenarios with minimal flavor violation in this context and how
they are subject to various empirical restrictions.
In one specific case, the relative strengths of the flavor-changing leptonic Higgs couplings
are determined mainly by the known quark mixing parameters and masses, and a branching fraction
\,${\cal B}(h\to\mu\tau)\sim1$\%\, is achievable without the couplings being incompatible
with the relevant constraints.
Upcoming data on the Higgs leptonic decays and searches for the \,$\mu\to e\gamma$\,
decay with improved precision can offer further tests on this scenario.
\end{abstract}

\maketitle

\section{Introduction\label{intro}}

The ongoing measurements on the 125\,GeV Higgs boson, $h$, at the Large Hadron Collider (LHC)
have begun to probe directly its Yukawa interactions with fermions\,\,\cite{atlas+cms,cms:h->2mu,
atlas:h->2mu,cms:h->mutau,cms:h->etau,atlas:h->mutau,lhc:t->hq}.
In particular, for the branching fractions of the standard decay modes of $h$, the ATLAS and
CMS experiments have so far come up with
\begin{eqnarray} \label{xBh2ff}
\frac{{\cal B}\big(h\to b\bar b\big)}{{\cal B}\big(h\to b\bar b\big)_{\textsc{sm}}}
\,=\, 0.70^{+0.29}_{-0.27} \mbox{\, \cite{atlas+cms}} \,, && ~~~~~
\frac{{\cal B}(h\to\tau^+\tau^-)}{{\cal B}(h\to\tau^+\tau^-)_{\textsc{sm}}^{}}
\,=\, 1.12^{+0.24}_{-0.22} \mbox{\, \cite{atlas+cms}} \,,
\nonumber \\
{\cal B}(h\to e^+e^-) \,<\, 0.0019 \mbox{\, \cite{cms:h->2mu}} \,, ~~ && \hspace{6ex}
{\cal B}(h\to\mu^+\mu^-) \,<\, 0.0015 \mbox{\, \cite{atlas:h->2mu}} \,,
\end{eqnarray}
where the upper limits in the second line are at 95\%\, confidence level (CL).
Overall, these data are still in harmony with the expectations of the standard model (SM).

However, there are also intriguing potential hints of physics beyond the SM in the Higgs
Yukawa couplings.
Especially, based on 19.7\,\,fb$^{-1}$ of Run-I data, CMS~\cite{cms:h->mutau}
has reported observing a slight excess of \,$h\to\mu^\pm\tau^\mp$\, events with
a significance of 2.4$\sigma$, which if interpreted as a signal implies
\begin{equation} \label{xBh2mt}
{\cal B}(h\to\mu\tau) \;=\; {\cal B}(h\to\mu^-\tau^+)+{\cal B}(h\to\mu^+\tau^-)
\;=\; \big(0.84^{+0.39}_{-0.37}\big)\% \,,
\end{equation}
but as a statistical fluctuation translates into the bound
\begin{equation} \label{xBh2mt'}
{\cal B}(h\to\mu\tau) \;<\; 1.51\%~~\mbox{at 95\% CL \cite{cms:h->mutau}} \,.
\end{equation}
Its ATLAS counterpart has a lower central value and bigger error,
\,${\cal B}(h\to\mu\tau)=(0.53\pm0.51)\%$\,
corresponding to \,${\cal B}(h\to\mu\tau)<1.43\%$\, at 95\% CL~\cite{atlas:h->mutau}.
Naively averaging the preceding CMS and ATLAS signal numbers, one would
get \,${\cal B}(h\to\mu\tau)=(0.73\pm0.31)\%$.\,
More recently, upon analyzing their Run-II data sample corresponding to 2.3 fb$^{-1}$,
CMS has found no excess and given the bound
\,${\cal B}(h \to \mu\tau)<1.20\%$\, at 95\% CL~\cite{CMS:2016qvi}.
This indicates that the analyzed integrated luminosity is not large enough to rule out
the Run-I excess and further analysis with more data is necessary to exclude or confirm it.
In contrast, although the observation of neutrino oscillation~\cite{pdg} suggests lepton flavor
violation, the SM contribution to lepton-flavor-violating Higgs decay via \mbox{$W$-boson} and
neutrino loops, with the neutrinos assumed to have mass, is highly suppressed due to both their
tiny masses and a Glashow-Iliopoulos-Maiani-like mechanism.
Therefore, the \,$h\to\mu\tau$\, excess events would constitute early evidence of new physics in
charged-lepton interactions if substantiated by future measurements.
On the other hand, searches for the $e\mu$ and $e\tau$ channels to date have produced
only the 95\%-CL bounds\,\,\cite{cms:h->etau}
\begin{eqnarray} \label{xBh2emu}
{\cal B}(h\to e\mu) \;<\; 0.036\,\% \,, ~~~~ ~~~
{\cal B}(h\to e\tau) \;<\; 0.70\,\%
\end{eqnarray}
from CMS and \,${\cal B}(h\to e\tau)<1.04\,\%$\, from ATLAS~\cite{atlas:h->mutau}.

In light of its low statistics, it is too soon to draw firm conclusions about
the tantalizing tentative indication of \,$h\to\mu\tau$\, in the present LHC data.
Nevertheless, in anticipation of upcoming measurements with improving precision, it is timely
to speculate on various aspects or implications of such a new-physics signal if it is discovered,
as has been done in very recent literature\,\,\cite{Dery:2013rta,h2mt,h2mt',h2mt'',He:2015rqa}.
In this paper, we assume that \,${\cal B}(h\to\mu\tau)\sim1\%$\, is realized in nature
and entertain the possibility that it arises from nonstandard effective Yukawa couplings
which may have some linkage to flavor-changing quark interactions beyond the SM.
For it is of interest to examine how the potential new physics responsible for
\,$h\to\mu\tau$\, may be subject to different constraints, including the current
nonobservation of Higgs-quark couplings deviating from their SM expectations.

To handle the flavor-violation pattern systematically without getting into model details,
we adopt the principle of so-called minimal flavor violation (MFV).
Motivated by the fact that the SM has been successful in describing the existing data on
flavor-changing neutral currents and $CP$ violation in the quark sector, the MFV hypothesis
presupposes that Yukawa couplings are the only sources for the breaking of flavor and $CP$
symmetries~\cite{mfv1,D'Ambrosio:2002ex}.
Unlike its straightforward application to quark processes, there is no unique way to
formulate leptonic MFV.
As flavor mixing among neutrinos has been empirically established\,\,\cite{pdg}, it is
attractive to formulate leptonic MFV by incorporating new ingredients that can
account for this fact~\cite{Cirigliano:2005ck}.
One could assume a\,\,minimal field content where only the SM lepton doublets and charged-lepton
singlets transform nontrivially under the flavor group, with lepton number violation and
neutrino masses coming from the dimension-five Weinberg operator~\cite{Cirigliano:2005ck}.
Less minimally, one could explicitly introduce right-handed
neutrinos~\cite{Cirigliano:2005ck}, or alternatively right-handed weak-SU(2)-triplet
fermions~\cite{He:2014efa}, which transform  nontrivially under an enlarged flavor
group and play an essential role in the seesaw mechanism to endow light neutrinos with
Majorana masses\,\,\cite{seesaw1,seesaw3}.
One could also introduce instead a\,\,weak-SU(2)-triplet of unflavored
scalars~\cite{He:2014efa,Gavela:2009cd} which participate in the seesaw
mechanism~\cite{seesaw2}.\footnote{Other aspects or scenarios of leptonic MFV have been
discussed in the literature\,\,\cite{Branco:2006hz,mlfv,He:2014fva,Grinstein:2006cg}.\smallskip}
Here we consider the SM expanded with the addition of three heavy right-handed neutrinos
as well as effective dimension-six operators conforming to the MFV criterion in both
the quark and lepton sectors.\footnote{\baselineskip=13pt%
A similar approach has been adopted in \cite{Lee:2015qra} to study some lepton-flavor-violating
processes that might occur as a consequence of the recently observed indications of anomalies in
rare \,$b\to s$\, decays.\smallskip}
To establish the link between the lepton and quark interactions beyond the SM, we consider
the implementation of MFV in a grand unified theory (GUT) framework\,\,\cite{Grinstein:2006cg}
with SU(5) as the unifying gauge
group\,\,\cite{Georgi:1974sy,Ellis:1979fg}.\footnote{\baselineskip=13pt%
A detailed analysis of the interplay between quark and lepton sectors in the framework of
a supersymmetric SU(5) GUT model with right-handed neutrinos can be found in~\cite{SUSY_GUT}.}
In this GUT scheme, there are mass relations between the SM charged leptons and down-type
quarks, and so we will deal with only the Higgs couplings to these fermions.

In the next section, we first briefly review the application of the MFV principle in a non-GUT
framework based on the SM somewhat enlarged with the inclusion of three right-handed
neutrinos which participate in the usual seesaw mechanism to generate light neutrino masses.
Subsequently, we introduce the effective dimension-six operators with MFV built-in that
can give rise to nonstandard flavor violation in Higgs decays, specifically the purely
fermionic channels\,\,$h\to f\bar f'$.\,
Then we look at constraints on the resulting flavor-changing Higgs couplings to quarks and
leptons, focusing on the former, as the leptonic case has been treated in detail in
Ref.\,\,\cite{He:2015rqa} which shows that the CMS \,$h\to\mu\tau$\, signal interpretation
can be explained under the MFV assumption provided that the right-handed
neutrinos couple to the Higgs in some nontrivial way.
In Section\,\,\ref{gutmfv}, we explore applying the MFV idea in the Georgi-Glashow SU(5)
GUT~\cite{Georgi:1974sy}, following the proposal of Ref.\,\,\cite{Grinstein:2006cg}.
As the flavor group is substantially smaller than in the non-GUT scheme, the number of
possible effective operators of interest becomes much larger.
Therefore, we will consider different scenarios involving one or more of the operators at
a time, subject to various experimental constraints.
We find that there are cases where the restrictions can be very severe if we
demand \,${\cal B}(h\to\mu\tau)\sim1$\%.\,
Nevertheless, we point out that there is an interesting scenario in which the flavor-changing
leptonic Higgs couplings depend mostly on the known quark mixing parameters and masses and
${\cal B}(h\to\mu\tau)$ at the percent level can occur in the parameter space allowed by other
empirical requirements.
Our analysis serves to illustrate that different possibilities in the GUT MFV context have
different implications for flavor-violating Higgs processes that may be testable in forthcoming
experiments.
We give our conclusions in Section\,\,\ref{conclusion}.
An appendix contains some extra information.

\section{Higgs fermionic decays with MFV\label{mfv}}

The renormalizable Lagrangian for fermion masses in the SM supplemented with three
right-handed Majorana neutrinos is
\begin{eqnarray} \label{Lm}
{\mathcal L}_{\rm m}^{} &\,=\,& -(Y_u)_{kl}^{}\,\overline{Q}_{k,L\,}^{}U_{l,R\,}^{} \tilde H
- (Y_d)_{kl}^{}\,\overline{Q}_{k,L\,}^{}D_{l,R\,}^{} H
- (Y_\nu)_{kl}^{}\,\overline{L}_{k,L\,}^{}\nu_{l,R\,}^{}\tilde H
- (Y_e)_{kl}^{}\,\overline{L}_{k,L\,}^{}E_{l,R\,}^{} H
\nonumber \\ && \!-~
\tfrac{1}{2} (M_\nu)_{kl}^{}\,\overline{(\nu_{k,R})\raisebox{1pt}{$^{\rm c}$}}\,\nu_{l,R}^{}
\;+\; {\rm H.c.} ~,
\end{eqnarray}
where summation over the family indices \,$k,l=1,2,3$\, is implicit, $Y_{u,d,\nu,e}$ denote
3$\times$3 matrices for the Yukawa couplings, $Q_{k,L}$ $(L_{k,L})$ is a left-handed
quark (lepton) doublet, $U_{l,R}$ and $D_{l,R\,}$ $\bigl(\nu_{l,R}^{}$ and $E_{l,R}\bigr)$
represent right-handed up- and down-type quarks (neutrinos and charged leptons), respectively,
$H$ stands for the Higgs doublet, \,$\tilde H=i\tau_2^{}H^*$\, with $\tau_2^{}$ being
the second Pauli matrix, $M_\nu$ is a 3$\times$3 matrix for the Majorana masses of $\nu_{l,R}$,
and the superscript of $(\nu_{k,R})^{\rm c}$ refers to charge conjugation.
We select the eigenvalues of $M_\nu$ to be much greater than the elements
of\,\,$v Y_\nu/\sqrt2$, so that the type-I seesaw mechanism becomes
operational~\cite{seesaw1}, leading to the light neutrinos' mass matrix
\,\mbox{$m_\nu^{}=-(v^2/2)\, Y_\nu^{}M_\nu^{-1}Y_\nu^{\textsc t}=U_{\textsc{pmns}\,}^{}
\hat m_{\nu\,}^{}U_{\textsc{pmns}}^{\textsc t}$},\,
which also involves the Higgs vacuum expectation value \,$v\simeq246$\,GeV,\,
the Pontecorvo-Maki-Nakagawa-Sakata (PMNS~\cite{pmns}) mixing matrix $U_{\textsc{pmns}}$ for
light neutrinos, and their eigenmasses $m_{1,2,3}^{}$ in
\,$\hat m_\nu^{}={\rm diag}\bigl(m_1^{},m_2^{},m_3^{}\bigr)$.\,
This suggests that~\cite{Casas:2001sr}
\begin{eqnarray} \label{Ynum}
Y_\nu^{} \;=\;
\frac{i\sqrt2}{v}\,U_{\textsc{pmns}\,}^{}\hat m^{1/2}_\nu OM_\nu^{1/2} \,,
\end{eqnarray}
where $O$ is in general a complex orthogonal matrix,
\,$OO^{\textsc t}=\openone\equiv{\rm diag}(1,1,1)$.\,

Hereafter, we suppose that $\nu_{k,R}$ are degenerate in mass, and so
\,$M_\nu={\mathcal M}\openone$.\,
The MFV hypothesis~\cite{D'Ambrosio:2002ex,Cirigliano:2005ck} then implies that
${\mathcal L}_{\rm m}$ is formally invariant under the global flavor symmetry group
\,${\mathcal G}_{\rm f}=G_q\times G_\ell$,\, where
\,$G_q^{}={\rm SU}(3)_Q\times{\rm SU}(3)_U\times{\rm SU}(3)_D$\, and
\,$G_\ell={\rm SU}(3)_L\times{\rm O}(3)_\nu\times{\rm SU}(3)_E$.\,
This entails that the above fermions are in the fundamental representations of their
respective flavor groups,
\begin{eqnarray}
Q_L^{} &\,\to\,& V_Q^{}Q_L^{} \,, ~~~~ ~~~ U_R^{} \,\to\, V_U^{}U_R^{} \,, ~~~~ ~~~
D_R^{} \,\to\, V_D^{}D_R^{} \,, ~~~~ ~~~ \nonumber \\
L_L^{} &\,\to\,&  V_L^{}L_L^{} \,, ~~~~~~~ \nu_R^{} \,\to\, {\mathcal O}_\nu^{}\nu_R^{} \,,
~~~~~~~ E_R^{} \,\to\, V_E^{}E_R^{} \,,
\end{eqnarray}
where \,$V_{Q,U,D,L,E}\in{\rm SU}(3)_{Q,U,D,L,E}$\, are special unitary matrices and
\,${\mathcal O}_\nu\in{\rm O}(3)_\nu$\, is an orthogonal real
matrix~\cite{D'Ambrosio:2002ex,Cirigliano:2005ck,Branco:2006hz}.
Moreover, the Yukawa couplings transform under ${\mathcal G}_{\rm f}$ in the spurion sense
according to
\begin{eqnarray}
Y_u^{} \,\to\, V_Q^{}Y_u^{}V^\dagger_U \,, ~~~~~ Y_d^{} \,\to\, V_Q^{}Y_d^{}V^\dagger_D \,, ~~~~~
Y_\nu^{} \,\to\, V_L^{}Y_\nu^{}{\mathcal O}_\nu^{\textsc t} \,, ~~~~~
Y_e^{} \,\to\, V_L^{}Y_e^{}V^\dagger_E \,.
\end{eqnarray}

To construct effective Lagrangians beyond the SM with MFV built-in, one inserts products of
the Yukawa matrices among the relevant fields to devise operators that are both
${\mathcal G}_{\rm f}$-invariant and singlet under the SM gauge
group~\cite{D'Ambrosio:2002ex,Cirigliano:2005ck}.
Of interest here are the combinations
\begin{eqnarray} \label{AB}
\textsf{A}_q^{} \,=\, Y_u^{}Y_u^\dagger \,, ~~~~~
\textsf{B}_q^{} \,=\, Y_d^{}Y_d^\dagger \,, ~~~~ ~~~
\textsf{A}_\ell^{} \,=\, Y_\nu^{}Y_\nu^\dagger \,, ~~~~~
\textsf{B}_\ell^{} \,=\, Y_e^{}Y_e^\dagger \,. ~~~~ ~~~
\end{eqnarray}
Given that the largest eigenvalues of $\textsf{A}_q$ and $\textsf{B}_q$ are
\,$y_t^2=2m_t^2/v^2\sim1$\, and \,$y_b^2=2m_b^2/v^2\sim3\times10^{-4}$, respectively,
at the mass scale \,$\mu\sim m_h^{}/2$,\, for our purposes we can devise objects containing
up to two powers of $\textsf{A}_q$ and drop contributions with $\textsf{B}_q$, as higher
powers of $\textsf{A}_q$ can be connected to lower ones by means of the Cayley-Hamilton
identity~\cite{Colangelo:2008qp}.
As for $\textsf{A}_\ell$, we assume that the right-handed neutrinos' mass is big enough,
\,${\mathcal M}\sim6\times10^{14}$\,GeV,\, to make the maximum eigenvalue of $\textsf{A}_\ell$
order\,\,1, which fulfills the perturbativity condition~\cite{He:2014fva,Colangelo:2008qp}.
Hence, as in the quark sector, we will keep terms up to order $\textsf{A}_\ell^2$ and
ignore those with $\textsf{B}_\ell$, whose elements are at most
\,$y_\tau^2=2m_\tau^2/v^2\sim10^{-4}$.\,
Accordingly, the relevant spurion building blocks are
\begin{eqnarray} \label{Delta}
\Delta_q^{} \,\,=\,\, \zeta^{}_1\openone + \zeta^{}_{2\,}\textsf{A}_q^{}
+ \zeta^{}_{4\,}\textsf{A}_q^2 \,, ~~~~~~~
\Delta_\ell^{} \,\,=\,\, \xi^{}_1\openone + \xi^{}_{2\,}\textsf{A}_\ell^{}
+ \xi^{}_{4\,}\textsf{A}_\ell^2 \,,
\end{eqnarray}
where in our model-independent approach $\zeta_{1,2,4}^{}$ and $\xi_{1,2,4}^{}$ are free
parameters expected to be at most of ${\mathcal O}(1)$ and with negligible imaginary
components~\cite{He:2014fva,Colangelo:2008qp}, so that one can make the approximations
\,$\Delta_q^\dagger=\Delta_q^{}$\, and \,$\Delta_\ell^\dagger=\Delta_\ell^{}$.\,

Thus, the desired ${\mathcal G}_{\rm f}$-invariant effective operators that are SM gauge
singlet and pertain to Higgs decays \,$h\to f\bar f'$\, into down-type fermions at tree
level are given by~\cite{Cirigliano:2005ck}\footnote{In this study, we do not address $h$
couplings to up-type quarks for the following reason. As the operator
\,$\big({\cal D}^\alpha\tilde H\big)\raisebox{1pt}{$^{\!\dagger\,}$}\overline{U}_R
Y_u^\dagger\Delta_q{\cal D}_\alpha Q_L$\,
with $\Delta_q$ from (\ref{Delta}) conserves flavor, others with
${\sf B}_q$, such as
\,$\big({\cal D}^\alpha\tilde H\big)\raisebox{1pt}{$^{\!\dagger\,}$}\overline{U}_R
Y_u^\dagger{\sf B}_q{\cal D}_\alpha Q_L$,\,
would be needed, but with only one Higgs doublet they are relatively suppressed by
the smallness of the ${\sf B}_q$ elements, which makes the present empirical
bounds~\cite{lhc:t->hq,Harnik:2012pb} on \,$t\to uh,ch$\,
and \,$h\to uc$\, not strong enough to offer meaningful constraints.}
\begin{eqnarray} \label{Lmfv}
{\cal L}_{\textsc{mfv}}^{} \,=\, \frac{O_{RL}^{}}{\Lambda^2} \,+\, {\rm H.c.} \,, ~~~~~
O_{RL}^{} \,=\, ({\cal D}^\alpha H)^{\dagger\,}\overline{D}_R^{}Y_d^\dagger\Delta_q^{}
{\cal D}_\alpha^{}Q_L^{} + ({\cal D}^\alpha H)^{\dagger\,}\overline{E}_R^{}Y_e^\dagger\Delta_\ell^{}
{\cal D}_\alpha^{}L_L^{} \,, ~~~
\end{eqnarray}
where the mass scale $\Lambda$ characterizes the underlying heavy new physics and
the covariant derivative
\,${\cal D}^\alpha=\partial^\alpha+(i g/2)\tau_a^{}W_a^\alpha+ig'Y'B^\alpha$\, acts on
\,$H,Q_L,L_L$\, with hypercharges \,$Y'=1/2,1/6,-1/2$,\, respectively, and
involves the usual SU(2)$_L\times{\rm U(1)}_Y$ gauge fields \,$W_a^\alpha$\, and $B^\alpha$,
their coupling constants $g$ and~$g'$, respectively, and Pauli matrices $\tau_a^{}$, with
\,$a=1,2,3$\, being summed over.
There are other dimension-six MFV operators involving $H$ and fermions, particularly
\begin{eqnarray} \label{others-q}
\begin{array}{ll}
i\bigl[H^{\dagger\,}{\cal D}_\alpha H-({\cal D}_\alpha H)^\dagger H \bigr]
\overline{Q}_L^{}\gamma^\alpha\Delta_{q1}^{}Q_L^{} \,,  &
g'\overline{D}_R^{}Y^\dagger_d\Delta_{q2\,}^{}\sigma_{\alpha\omega}^{}
H^\dagger Q_L^{}B^{\alpha\omega} \,,
\vspace{1ex} \\
i\bigl[ H^\dagger\tau_{a\,}^{}{\cal D}_\alpha H
- ({\cal D}_\alpha H)^\dagger\tau_a^{}H \bigr] \overline{Q}_L^{}\gamma^\alpha
\Delta_{q3\,}^{}\tau_a^{}Q_L^{} \;, ~~~ ~~~~ &
g_{\,}\overline{D}_R^{}Y^\dagger_d\Delta_{q4\,}^{}\sigma_{\alpha\omega}^{}
H^\dagger\tau_a^{}Q_L^{}W_a^{\alpha\omega}
\end{array}
\end{eqnarray}
in the quark sector and
\begin{eqnarray} \label{others-l}
\begin{array}{ll}
i\bigl[H^{\dagger\,}{\cal D}_\alpha H-({\cal D}_\alpha H)^\dagger H \bigr]
\overline{L}_L^{}\gamma^\alpha\Delta_{\ell1}^{}L_L^{} \,,  &
g'\overline{E}_R^{}Y^\dagger_e\Delta_{\ell2\,}^{}\sigma_{\alpha\omega}^{}
H^\dagger L_L^{}B^{\alpha\omega} \,,
\vspace{1ex} \\
i\bigl[ H^\dagger\tau_{a\,}^{}{\cal D}_\alpha H
- ({\cal D}_\alpha H)^\dagger\tau_a^{}H \bigr] \overline{L}_L^{}\gamma^\alpha
\Delta_{\ell3\,}^{}\tau_a^{}L_L^{} \;, ~~~ ~~~~ &
g_{\,}\overline{E}_R^{}Y_e^\dagger\Delta_{\ell4\,}^{}\sigma_{\alpha\omega}^{}
H^\dagger\tau_a^{}L_L^{}W_a^{\alpha\omega}
\end{array}
\end{eqnarray}
in the lepton sector, where $\Delta_{qn}$ and $\Delta_{\ell n}$ are the same in form as
$\Delta_q$ and $\Delta_\ell$, respectively, except they have their own coefficients
$\zeta_r^{}$ and $\xi_r^{}$, but these operators do not induce \,$h\to f\bar f'$\, at tree level.
In the literature the operators
\,$H^\dagger H_{\,}\overline{D}_R^{}Y_d^\dagger\Delta_q^{}H^\dagger Q_L^{}$\,
and \,$H^\dagger H_{\,}\overline{E}_R^{}Y_e^\dagger\Delta_\ell^{~}H^\dagger L_L^{}$\,
are also often considered ({\it e.g.},\,\,\cite{Dery:2013rta}), but they can be shown
using the equations of motion for SM fields to be related to $O_{RL}^{}$ and
the other operators above~\cite{Grzadkowski:2010es}.\footnote{This was explicitly done for
the leptonic operators in~\cite{He:2015rqa}.\medskip}

It is worth remarking that there are relations among $\Delta_q$ and $\Delta_{qn}$
above (among their respective sets of coefficients $\zeta_r^{}$) which are fixed within
a given model, but such relations are generally different in a different model.
As a consequence, stringent bounds on processes induced by one or more of the quark operators
in Eqs.\,\,(\ref{Lmfv}) and (\ref{others-q}) may not necessarily apply to the others,
depending on the underlying new-physics model.
Similar statements can be made regarding $\Delta_\ell$,\,\,$\Delta_{\ell n}$, and the lepton
operators in Eqs.\,\,(\ref{Lmfv}) and (\ref{others-l}).\footnote{\baselineskip=13pt%
The high degree of model dependency in the relationships among the $\Delta$s belonging to
the different operators is well illustrated by the results of the papers in
\cite{h2mt,h2mt',h2mt''} which address \,$h\to\mu\tau$\, in the contexts of various scenarios.
Particularly, there are models~\cite{h2mt'} in which \,${\cal B}(h\to\mu\tau)\sim1$\%\, is
achievable from tree-level contributions without much hindrance from the strict experimental
requirements on \,$\ell\to\ell'\gamma$\, transitions, including lepton $g-2$, which arise
from one-loop diagrams.
In some other models~\cite{h2mt''} all these processes only occur at the loop level and the limiting
impact of the \,$\ell\to\ell'\gamma$\, restrictions on \,$h\to\mu\tau$ is considerable.
It follows that one cannot make definite predictions for \,$\ell\to\ell'\gamma$\, in
a\,\,model-independent way based on the input from \,$h\to\mu\tau$.}
For these reasons, in our model-independent analysis on the contributions of $O_{RL}^{}$ to
\,$h\to f\bar f'$\, we will not deal with constraints on the operators in
Eqs.\,\,(\ref{others-q}) and\,\,(\ref{others-l}).
Our results would then implicitly pertain to scenarios in which such constraints do not
significantly affect the predictions for \,$h\to f\bar f'$.\,

In view of $O_{RL}^{}$ in Eq.\,(\ref{Lmfv}) which is invariant under the flavor symmetry
${\cal G}_{\rm f}$, it is convenient to rotate the fields and work in the basis
where $Y_{d,e}$ are diagonal,
\begin{eqnarray} \label{YdYe}
Y_d^{} \,=\, {\rm diag}\bigl(y_d^{},y_s^{},y_b^{}\bigr) \,, ~~~~ ~~~
Y_e^{} \,=\, {\rm diag}\bigl(y_e^{},y_\mu^{},y_\tau^{}\bigr) \,, ~~~~ ~~~
y_f^{} \,=\, \sqrt2\,m_f^{}/v \,,
\end{eqnarray}
and $U_k$, $D_k$, $\tilde\nu_{k,L}$, $\nu_{k,R}$, and $E_k$ refer to the mass eigenstates.
Explicitly, \,$(U_1,U_2,U_3)=(u,c,t)$,\, $(D_1,D_2,D_3)=(d,s,b)$,\,
and \,$(E_1,E_2,E_3)=(e,\mu,\tau)$.\,
Accordingly,
\begin{eqnarray}
Q_{k,L}^{} &=& \left(\!\begin{array}{c} (V^\dagger_{\textsc{ckm}})_{kl\,}^{}U_{l,L}^{} \\
D_{k,L}^{} \end{array}\!\right) , ~~~~ ~
L_{k,L}^{} \,= \left(\!\begin{array}{c} (U_{\textsc{pmns}})_{kl}^{}\,
\tilde\nu_{l,L}^{} \vspace{2pt} \\ E_{k,L}^{} \end{array}\!\right) , ~~~~ ~
Y_u \,=\, V^\dagger_{\textsc{ckm}}\,{\rm diag}\bigl(y_u^{},y_c^{},y_t^{}\bigr) \,,
\nonumber \\
\textsf{A}_q^{} &=& V^\dagger_{\textsc{ckm}}\,
{\rm diag}\bigl(y_u^2,y_c^2,y_t^2\bigr)\, V_{\textsc{ckm}}^{} \,, ~~~~ ~~~~ ~~~
\textsf{A}_\ell^{} \,=\, \frac{2\mathcal M}{v^2}\, U_{\textsc{pmns}\,}^{} \hat m^{1/2}_\nu
O O^\dagger\hat m^{1/2}_\nu U_{\textsc{pmns}}^\dagger \,,
\nonumber \\ \label{BqBl}
\textsf{B}_q^{} &=& {\rm diag}\bigl(y_d^2,y_s^2,y_b^2\bigr) \,, \hspace{8em}
\textsf{B}_\ell^{} \,=\, {\rm diag}\bigl(y_e^2,y_\mu^2,y_\tau^2\bigr) \,,
\end{eqnarray}
where $V_{\textsc{ckm}}$ is the Cabibbo-Kobayashi-Maskawa (CKM) quark mixing matrix.

Now, we express the effective Lagrangian describing \,$h\to f\bar f'$\, as
\begin{eqnarray}
{\cal L}_{h f\bar f'} \;=\; -\overline{f}\big({\cal Y}_{f'f}^*P_L^{}+{\cal Y}_{ff'}^{}P_R^{}
\big)f_{\,}'h \,,
\end{eqnarray}
where ${\cal Y}_{ff',f'f}$ are the Yukawa couplings, which are generally complex,
and \,$P_{L,R}=(1 \mp\gamma_5)/2$\, are chirality projection operators.\,
This leads to the decay rate
\begin{eqnarray}
\Gamma_{h\to f\bar f'}^{} \,=\, \frac{m_h^{}}{16\pi} \Big(
\big|{\cal Y}_{f'f}^{}\big|\raisebox{1pt}{$^2$} +
\big|{\cal Y}_{ff'}^{}\big|\raisebox{1pt}{$^2$}\Bigr) \,,
\end{eqnarray}
where the fermion masses have been neglected compared to $m_h^{}$.
Thus, from Eq.\,(\ref{Lmfv}), which contributes to both flavor-conserving and -violating
transitions, we find for \,$h\to D_k^{}\bar D_l^{},E_k^-E_l^+$\,
\begin{eqnarray} \label{YDkDl}
{\cal Y}_{D_kD_l}^{} &=& {\cal Y}_{D_kD_l}^{\textsc{sm}}
\,-\, \frac{m_{D_l}^{}m_h^2}{2\Lambda^2v}\,(\Delta_q)_{kl}^{} \,,
\\ \label{YEkEl}
{\cal Y}_{E_kE_l}^{} &=& \delta_{kl}^{}\,{\cal Y}_{E_kE_k}^{\textsc{sm}}
\,-\, \frac{m_{E_l}^{}m_h^2}{2\Lambda^2v}\,(\Delta_\ell)_{kl}^{} \,,
\end{eqnarray}
where we have included the SM contributions, which are separated from the $\Delta_{q,\ell}$
terms and can be flavor violating only in the quark case due to loop effects,
and \,${\cal Y}_{ff}^{\textsc{sm}}=m_f^{}/v$\, at tree level.
Since approximately \,$\Delta_{q,\ell}^{}=\Delta_{q,\ell}^\dagger$,\, it follows that
in our MFV scenario \,$|{\cal Y}_{ff'}|\gg|{\cal Y}_{f'f}|$\, for
\,$ff'=ds,db,sb,e\mu,e\tau,\mu\tau$\, and ${\cal Y}_{ff}$ are real.

For ${\cal Y}_{ds,db,sb}$, it is instructive to see how they compare to each other
in the presence of $\Delta_q$.
In terms of the Wolfenstein parameters $(\lambda,A,\rho,\eta)$, the matrices ${\sf A}_q$ and
${\sf A}_q^2$ in $\Delta_q$ are given by
\begin{eqnarray} \label{Aq}
{\sf A}_q^{} \,\simeq \left(\begin{array}{ccc} \lambda^6 A^2\big[(1-\rho)^2+\eta^2\big] &
~~ \mbox{$-\lambda^5$}A^2(1-\rho+i\eta) ~~ & \lambda^3 A (1-\rho+i\eta) \vspace{3pt} \\
-\lambda^5 A^2(1-\rho-i\eta) & \lambda^4 A^2 & -\lambda^2 A \vspace{3pt} \\
\lambda^3 A (1-\rho-i\eta) & -\lambda^2 A & 1 \end{array}\right)
\simeq\, {\sf A}_q^2
\end{eqnarray}
to the lowest nonzero order in \,$\lambda\simeq0.23$\, for each component, as
\,$y_u^2\ll y_c^2\sim1.4\times10^{-5}\sim2\lambda^8$\, and  \,$y_t^{}\sim1$\,
at the renormalization scale \,$\mu\sim m_h^{}/2$.\,
If the $\Delta_q$ part of ${\cal Y}_{D_kD_l}$ for \,$k\neq l$\, is dominant,
we then arrive at the ratio
\begin{eqnarray} \label{Yqq'ratio}
|{\cal Y}_{ds}| : |{\cal Y}_{db}| : |{\cal Y}_{sb}| \;\simeq\;
\lambda^3A|1-\rho+i\eta|m_s^{}:\lambda|1-\rho+i\eta|m_b^{}:m_b^{} \;=\; 0.00016 : 0.21 : 1 \,,
\end{eqnarray}
the numbers having been calculated with the central values of the Wolfenstein parameters
from Ref.\,\,\cite{ckmfit}\footnote{Explicitly, \,$\lambda=0.22543$, \,$A=0.823$,
\,$\rho\simeq0.1536$,\, and \,$\eta\simeq0.3632$.}
as well as \,$m_s^{}=57$\,MeV\, and \,$m_b^{}=3.0$\,GeV\, at \,$\mu\sim m_h^{}/2$.\,

\begin{figure}[t]
\includegraphics[width=7cm]{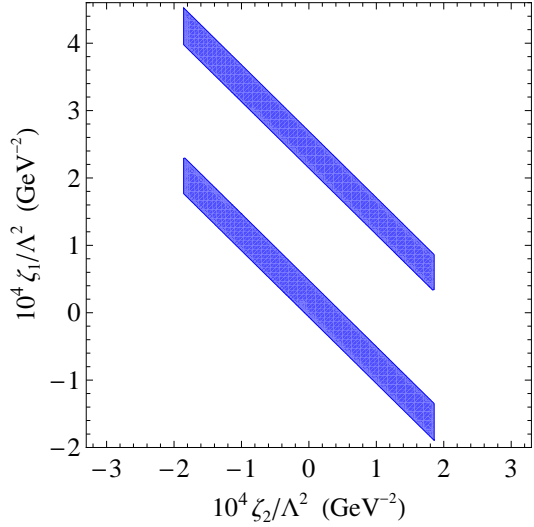} \vspace{-1ex}
\caption{Regions of $\zeta_1^{}/\Lambda^2$ and $\zeta_2^{}/\Lambda^2$ for \,$\zeta_4^{}=0$\,
which fulfill the experimental constraints in Eqs.\,(\ref{|Yqq'|limits})-(\ref{|Ybb|limit}).
The $\zeta_2^{}/\Lambda^2$ range is determined by
\,$|{\cal Y}_{db}|^2<2.3\times10^{-8}$\, from Eq.\,(\ref{|Yqq'|limits}).\label{z1z2-qq}}
\end{figure}

The SM coupling ${\cal Y}_{D_kD_l}^{\textsc{sm}}$ with \,$k\neq l$\, arises from one-loop
diagrams with the $W$ boson and up-type quarks in the loops.
Numerically, we employ the formulas available from Ref.\,\,\cite{Dedes:2003kp} to obtain
\,${\cal Y}_{ds}^{\textsc{sm}}=(7.2+3.1i)\times10^{-10}$,
\,${\cal Y}_{db}^{\textsc{sm}}=-(9.2+3.8i)\times10^{-7}$,
\,${\cal Y}_{sb}^{\textsc{sm}}=(4.7-0.1i)\times10^{-6}$,\,
and relatively much smaller $\big|{\cal Y}_{sd,bd,bs}^{\textsc{sm}}\big|$.
These SM predictions are, as expected, consistent with the ratio in
Eq.\,(\ref{Yqq'ratio}), but still lie very well within the indirect bounds
inferred from the data on $K$-$\bar K$, $B_d$-$\bar B_d$, and $B_s$-$\bar B_s$
oscillations, namely~\cite{Harnik:2012pb}
\begin{eqnarray} \label{|Yqq'|limits}
-5.9\times10^{-10} \,<\, {\rm Re}\big({\cal Y}_{ds,\,sd}^2\big) \,<\, 5.6\times10^{-10} , ~
&& ~~~
\big|{\rm Re}\big({\cal Y}_{ds}^*{\cal Y}_{sd}^{}\big)\big| \,<\, 5.6\times10^{-11} ,
\nonumber \\
-2.9\times10^{-12} \,<\, {\rm Im}\big({\cal Y}_{ds,\,sd}^2\big) \,<\, 1.6\times10^{-12} , ~
&& ~~~
\mbox{$-1.4\times10^{-13}$} \,<\, {\rm Im}\big({\cal Y}_{ds}^*{\cal Y}_{sd}^{}\big)
\,<\, 2.8\times10^{-13} ,
\nonumber \\
|{\cal Y}_{db,bd}^{}|^2 \,<\, 2.3\times10^{-8} \,, \hspace{7em} && \hspace{7ex}
|{\cal Y}_{db}^{~~}{\cal Y}_{bd}^{}\big| \,<\, 3.3\times10^{-9} \,,
\nonumber \\
|{\cal Y}_{sb,bs}^{}|^2 \,<\, 1.8\times10^{-6} \,, \hspace{7em} && \hspace{7ex}
|{\cal Y}_{sb}^{~~}{\cal Y}_{bs}^{}| \,<\, 2.5\times10^{-7} \,.
\end{eqnarray}
Hence there is ample room for new physics to saturate one or more of these limits.
Before examining how the ${\cal L}_{\textsc{mfv}}$ contributions may do so, we need to take
into account also the \,$h\to b\bar b$\, measurement quoted in Eq.\,(\ref{xBh2ff}).
Thus, based on the 90\%-CL range of this number in view of its currently sizable error,
we may impose
\begin{eqnarray} \label{|Ybb|limit}
0.4 \,<\, |{\cal Y}_{bb}^{}/{\cal Y}_{bb}^{\textsc{sm}}|^2 \,<\, 1.1 \,,
\end{eqnarray}
where \,${\cal Y}_{bb}^{\textsc{sm}}\simeq0.0125$\, from the central values of the SM Higgs
total width \,$\Gamma_h^{\textsc{sm}}=4.08\;$MeV\, and
\,${\cal B}\big(h\to b\bar b\big){}_{\textsc{sm}}^{}=0.575$\,
determined in Ref.\,\,\cite{lhctwiki} for \,$m_h^{}=125.1\;$GeV~\cite{pdg}.
Upon applying the preceding constraints to Eq.\,(\ref{YDkDl}), we learn that
\,$|{\cal Y}_{db}|^2<2.3\times10^{-8}$\, in Eq.\,(\ref{|Yqq'|limits})
and the one in Eq.\,(\ref{|Ybb|limit}) are the most consequential and that the former can be
saturated if at least both the $\zeta_1^{}$ and $\zeta_2^{}$, or $\zeta_4^{}$, terms in
$\Delta_q$ are nonzero.
We illustrate this in Fig.\,\ref{z1z2-qq} for \,$\zeta_4^{}=0$,\, where
the $\zeta_2^{}/\Lambda^2$ limits of the (blue) shaded areas are fixed by the just mentioned
$|{\cal Y}_{db}|$ bound and the $\zeta_1^{}/\Lambda^2$ values in these areas ensure that
Eq.\,(\ref{|Ybb|limit}) is satisfied.
Interchanging the roles of $\zeta_2^{}$ and $\zeta_4^{}$ would lead to an almost identical plot.
If \,$|\zeta_{1,2}^{}|\sim1$,\, these results imply a fairly weak lower-limit on the MFV
scale $\Lambda$ of around \,50\,\,GeV.\,

For the leptonic Yukawa couplings, ${\cal Y}_{E_kE_l}$ in Eq.\,(\ref{YEkEl}), the situation
is different and not unique because the specific values and relative sizes of the elements
of ${\sf A}_\ell$ in $\Delta_\ell$ can vary greatly~\cite{He:2015rqa}.
In our MFV scenario with the type-I seesaw, this depends on the choices of the right-handed
neutrinos' mass $\cal M$ and the orthogonal matrix $O$ as well as on whether the light
neutrinos' mass spectrum $(m_1,m_2,m_3)$ has a normal hierarchy (NH) or an inverted one (IH).

For instance, if $O$ is real,
\,${\sf A}_\ell=\big(2\mathcal M/v^2\big)U_{\textsc{pmns}\,}^{}
\hat m_{\nu\,}^{}U_{\textsc{pmns}}^\dagger$\,
from Eq.\,(\ref{BqBl}), and using the central values of neutrino mixing parameters from
a recent fit to global neutrino data~\cite{nudata} we find in the NH case with \,$m_1=0$\,
\begin{eqnarray} \label{AlNH}
{\sf A}_\ell^{} \,\simeq \frac{10^{-15}\mathcal{M}}{\rm GeV} \left(\begin{array}{ccc} 0.12 &
0.19+0.12i & 0.01+0.14i \vspace{2pt} \\ 0.19-0.12i & 0.82 & 0.7-0.02i \vspace{2pt} \\
0.01-0.14i & ~0.70+0.02i~ & 0.98 \end{array}\right) .
\end{eqnarray}
Incorporating this and selecting \,$\xi_4^{}=0$\, in $\Delta_\ell$ to be employed in
Eq.\,(\ref{YEkEl}), we then arrive at
\,$|{\cal Y}_{e\mu}|:|{\cal Y}_{e\tau}|:|{\cal Y}_{\mu\tau}|=
|({\sf A}_\ell)_{12}^{}|m_\mu^{}:|({\sf A}_\ell)_{13}^{}|m_\tau^{}:
|({\sf A}_\ell)_{23}^{}|m_\tau^{}\simeq0.019:0.19:1$.\,
Interchanging the roles of $\xi_2^{}$ and $\xi_4^{}$ would modify the ratio to
\,$0.013:0.21:1$.\,
In the IH case with \,$m_3=0$,\, the corresponding numbers are roughly about the same.
These results for the Yukawas in the real-$O$ case turn out to be incompatible with
the following experimental constraints on the Yukawa couplings if we demand
\,${\cal B}(h\to\mu\tau)\sim1$\%\, as CMS suggested, but with $O$ being
complex instead it is possible to satisfy all of these requirements~\cite{He:2015rqa}.

For the first set of constraints, the direct-search limits in Eqs.\,(\ref{xBh2mt'}) and
(\ref{xBh2emu}) translate into~\cite{cms:h->etau}
\begin{align} \label{|Yll'|limit}
\sqrt{|{\cal Y}_{e\mu}^{}|^2+|{\cal Y}_{\mu e}^{}|^2} \,<\, 5.43\times10^{-4} \,, ~~~~ ~~~
\sqrt{|{\cal Y}_{e\tau}^{}|^2+|{\cal Y}_{\tau e}^{}|^2} \,<\, 2.41\times10^{-3} \,,
\end{align}
and \,$\sqrt{|{\cal Y}_{\mu\tau}|^2+|{\cal Y}_{\tau\mu}|^2}<3.6\times10^{-3}$\,
under the no-signal assumption, while Eq.\,(\ref{xBh2mt}) for the \,$h\to\mu\tau$\,
signal interpretation implies
\begin{eqnarray} \label{|Ymt|limit}
2.0\times10^{-3} \,<\, \sqrt{\big|{\cal Y}_{\tau\mu}^{}\big|\raisebox{1pt}{$^2$} +
\big|{\cal Y}_{\mu\tau}^{}\big|\raisebox{1pt}{$^2$}} \,<\, 3.3\times10^{-3} \,.
\end{eqnarray}
Additionally, the latest experimental bound
\,${\cal B}(\mu\to e\gamma)<4.2\times10^{-13}$\, at 90\% CL~\cite{meg}
on the loop-induced decay \,$\mu\to e\gamma$\, can offer a complementary, albeit indirect,
restraint~\cite{Dery:2013rta,Harnik:2012pb,Goudelis:2011un} on different
couplings simultaneously~\cite{He:2015rqa}
\begin{eqnarray} \label{m2eglimit}
\sqrt{\big|\big({\cal Y}_{\mu\mu}^{}+r_\mu^{}\big){\cal Y}_{\mu e}^{} +
9.19\,{\cal Y}_{\mu\tau\,}^{}{\cal Y}_{\tau e}^{}\big|\raisebox{1pt}{$^2$} +
\big|\big({\cal Y}_{\mu\mu}^{}+r_\mu^{}\big){\cal Y}_{e\mu}^{} +
9.19\,{\cal Y}_{e\tau\,}^{}{\cal Y}_{\tau\mu}^{}\big|\raisebox{1pt}{$^2$}}
\,<\, 4.4\times10^{-7} \,, ~~~~
\end{eqnarray}
with \,$r_\mu^{}=0.29$\, \cite{Harnik:2012pb}.
This could be stricter especially on ${\cal Y}_{e\mu,\mu e}$ than its direct counterpart in
Eq.\,(\ref{|Yll'|limit}) if destructive interference with other potential new physics effects
is absent.
Compared to Eqs.\,(\ref{|Yll'|limit})-(\ref{m2eglimit}), the indirect
limits~\cite{Harnik:2012pb} from the data on \,$\tau\to e\gamma,\mu\gamma$\, and leptonic
anomalous magnetic and electric dipole moments are not competitive for our MFV cases.
Finally, the \,$h\to\mu^+\mu^-,\tau^+\tau^-$\, measurements quoted in Eq.\,(\ref{xBh2ff})
are also relevant and may be translated into
\begin{eqnarray} \label{|Yll|limit}
\big|{\cal Y}_{\mu\mu}^{}/{\cal Y}_{\mu\mu}^{\textsc{sm}}\big|^2 \,<\, 5 \,, ~~~~~
0.9 \,<\, \big|{\cal Y}_{\tau\tau}^{}/{\cal Y}_{\tau\tau}^{\textsc{sm}}\big|^2 \,<\, 1.3 \,,
\end{eqnarray}
where \,${\cal Y}_{\mu\mu}^{\textsc{sm}}\simeq4.24\times10^{-4}$\, and
\,${\cal Y}_{\tau\tau}^{\textsc{sm}}\simeq7.19\times10^{-3}$\, from
\,${\cal B}(h\to\mu^+\mu^-)_{\textsc{sm}}^{}=2.19\times10^{-4}$\, and
\,${\cal B}(h\to\tau^+\tau^-)_{\textsc{sm}}^{}=6.30$\%\, supplied by Ref.\,\,\cite{lhctwiki}.

As pointed out in Ref.\,\,\cite{He:2015rqa}, the aforementioned leptonic MFV scenario with
the $O$ matrix in ${\textsf A}_\ell$ being real is unable to accommodate the preceding constraints,
especially Eqs.\,(\ref{|Ymt|limit}) and\,\,(\ref{m2eglimit}), even with the $\xi_{1,2,4}^{}$
terms in $\Delta_\ell$ contributing at the same time.
Rather, it is necessary to adopt a less simple structure of ${\textsf A}_\ell$ with $O$ being
complex, which can supply extra free parameters to achieve the desired results, one of them
being \,$|{\cal Y}_{e\mu}/{\cal Y}_{\mu\tau}|\,\mbox{\footnotesize$\lesssim$}\,10^{-3}$.\,
This possibility was already explored in Ref.\,\,\cite{He:2015rqa} and therefore will
not be analyzed further here.

\section{Higgs fermionic decays in GUT with MFV\label{gutmfv}}

In the Georgi-Glashow grand unification based on the SU(5) gauge
group~\cite{Georgi:1974sy}\footnote{For a review see, {\it e.g.}, \cite{Langacker:1980js}.}
the conjugate of the right-handed down-type quark, $(D_{k,R})^{\rm c}$, and the left-handed
lepton doublet, $L_{k,L}$, appear in the {\boldmath\small$\bar 5$} representations $\psi_k^{}$,
whereas the left-handed quark doublets, $Q_{k,L}$, and the conjugates of the right-handed
up-type quark and charged lepton, $(U_{k,R})^{\rm c}$ and $(E_{k,R})^{\rm c}$, belong to
the {\bf\small10} representations $\chi_k^{}$.
With three SU(5)-singlet right-handed neutrinos being included in the theory, the Lagrangian
for fermion masses is~\cite{Grinstein:2006cg,Ellis:1979fg}
\begin{eqnarray}
{\cal L}_{\rm m}^{\textsc{gut}} &=&
(\lambda_5)_{kl}^{}\, \psi_k^{\textsc t}\chi_{l\,}^{}{\sf H}_5^*
+ (\lambda_{10})_{kl}^{}\, \chi_k^{\textsc t}\chi_{l\,}^{}{\sf H}_5^{} \,+\,
\frac{(\lambda_5')_{kl}^{}}{M_{\textsc p}}\,\psi_k^{\textsc t}\Sigma_{24}^{~~\;}\chi_{l\,}^{}{\sf H}_5^*
\nonumber \\ && \! +\;
(\lambda_1)_{kl}^{}\,\nu_{k,R}^{\textsc t~~}\psi_{l\,}^{}{\sf H}_5^{}
\,-\, \frac{(M_\nu)_{kl}^{}}{2}\, \nu_{k,R}^{\textsc t~~}\nu_{l,R}^{} \;+\; {\rm H.c.} \,,
\end{eqnarray}
where SU(5) indices have been dropped, ${\sf H}_5$ and $\Sigma_{24}$ are Higgs fields in
the {\bf\small5} and {\bf\small24} of SU(5), and compared to the GUT scale
the Planck scale \,$M_{\textsc p}\gg M_{\textsc{gut}}$.\,
Since ${\cal L}_{\rm m}^{\textsc{gut}}$ contains ${\cal L}_{\rm m}$ for the SM
plus 3 degenerate right-handed neutrinos, the Yukawa
couplings in these Lagrangians satisfy the relations~\cite{Grinstein:2006cg,Ellis:1979fg}
\begin{eqnarray} \label{Ygut}
Y_u^\dagger \,\propto\, \lambda_{10}^{} \,, ~~~~~
Y_d^\dagger \,\propto\, \lambda_5^{}+\epsilon\lambda_5' \,, ~~~~~
Y_e^* \,\propto\, \lambda_5^{}-\tfrac{3}{2}\,\epsilon\lambda_5' \,, ~~~~~
Y_\nu^\dagger \,=\, \lambda_1^{} \,,
\end{eqnarray}
where \,$\epsilon=M_{\textsc{gut}}/M_{\textsc p}\ll1$.\,
Evidently, in the absence of the dimension-five nonrenormalizable $\lambda_5'$ term in
${\cal L}_{\rm m}^{\textsc{gut}}$ the down-type Yukawas would be related
by \,$Y_d^{}\propto Y_e^{\textsc t}$\, which is inconsistent with the experimental
masses~\cite{Ellis:1979fg}.
In this work, we do not include the corresponding term for the up-type quark sector,
\,$(\lambda_{10}')_{kl}^{}\,\chi_k^{\textsc t}\Sigma_{24}^{~~\;}
\chi_{l\,}^{}{\sf H}_5^{}/M_{\textsc p}^{}$\, \cite{Grinstein:2006cg},
which could significantly correct the up-quark mass, but does not lead to any quark-lepton
mass relations.

The application of the MFV principle in this GUT context entails that under the global flavor
symmetry group
\,${\mathcal G}_{\rm f}^{\textsc{gut}}=
{\rm SU}(3)_{\bar 5}\times{\rm SU}(3)_{10}\times{\rm O}(3)_1$\,
the fermion fields and Yukawa spurions in ${\cal L}_{\rm m}^{\textsc{gut}}$ transform
as~\cite{Grinstein:2006cg}
\begin{eqnarray}
\psi &\to& V_{\bar 5\,}^{}\psi \,, \hspace{5em}\,
\chi \,\to\, V_{10\,}^{}\chi \,, \hspace{4em}\, \nu_R^{} \,\to\, {\mathcal O}_1^{}\nu_R^{} \,,
\nonumber \\
\lambda_5^{\scriptscriptstyle(\prime)} &\to&
V_{\bar 5}^*\lambda_5^{\scriptscriptstyle(\prime)} V_{10}^\dagger \,, ~~~~ ~~~
\lambda_{10}^{\scriptscriptstyle(\prime)} \,\to\,
V_{10}^*\lambda_{10}^{\scriptscriptstyle(\prime)}V_{10}^\dagger \,, ~~~~ ~~~
\lambda_1^{} \,\to\, {\mathcal O}_1^{}\lambda_1^{}V_{\bar 5}^\dagger \,,
\end{eqnarray}
where we have assumed again that the right-handed neutrinos are degenerate,
\,$V_{\bar 5,10}\in{\rm SU(3)}_{\bar 5,10}$,\, and \,${\mathcal O}_1\in{\rm O}(3)_1^{}$.\,
It follows that the flavor transformation properties of the fermions
and Yukawa coupling matrices in ${\cal L}_{\rm m}$ are
\begin{eqnarray} \label{f'=Vf}
Q_L^{} &\to& V_{10}^{~~} Q_L^{} \,, ~~~~ ~~~~ ~ U_R^{} \,\to\, V_{10}^* U_R^{} \,, ~~~~ ~~~~ ~
D_R^{} \,\to\, V_{\bar 5}^* D_R^{} \,,
\nonumber \\
L_L^{} &\to& V_{\bar 5}^{~}L_L^{} \,, ~~~~ ~~~~ ~~ E_R^{} \,\to\, V_{10}^* E_R^{} \,,
\nonumber \\
Y_u^{} &\to& V_{10}^{~~} Y_u^{~} V_{10}^{\textsc t} \,, ~~~~~ ~
Y_d^{} \;\to\; V_{10}^{~~} Y_d^{~} V_{\bar 5}^{\textsc t} \,, ~~~~ ~~
Y_e^{} \;\to\; V_{\bar 5}^{~} Y_e^{~} V_{10}^{\textsc t} \,, ~~~~~
\nonumber \\
Y_\nu^{} &\to& V_{\bar 5}^{~} Y_\nu^{~} {\cal O}_1^{\textsc t} \,.
\end{eqnarray}
As in the non-GUT scheme treated in the previous section, one can then put together
the spurion building blocks
\,$\Delta_q^{}=\zeta_1^{}\openone+\zeta_{2\,}^{}{\sf A}_q^{}+\zeta_{4\,}^{}{\sf A}_q^2$\,
and
\,$\Delta_\ell^{}=\xi_1^{}\openone+\xi_{2\,}^{}{\sf A}_\ell^{}+\xi_{4\,}^{}{\sf A}_\ell^2$,\,
after dropping contributions involving products of down-type Yukawas, which have more
suppressed elements.\footnote{Like before, we have assumed that the right-handed neutrinos'
mass \,${\cal M}\sim6\times10^{14}$\,GeV,\, so that the biggest eigenvalue of ${\sf A}_\ell$
is around one.
Otherwise, if \,${\cal M}\ll10^{14}$\,GeV,\, the flavor-violating impact of $\Delta_\ell$ would
decrease accordingly.}

In analogy to the non-GUT scenario, the effective operators of interest constructed out of
the spurions and SM fields need to be invariant under both
${\mathcal G}_{\rm f}^{\textsc{gut}}$ and the SM gauge group.
However, since ${\mathcal G}_{\rm f}^{\textsc{gut}}$ is significantly smaller than
${\mathcal G}_{\rm f}$, in the GUT MFV framework there are many more ways to arrange
flavor-symmetry-breaking objects for the operators~\cite{Grinstein:2006cg}.
It is straightforward to see that those pertaining to Higgs decays into down-type fermions
at tree level are given by
\begin{eqnarray} \label{Lgutmfv0}
{\cal L}_{\textsc{mfv}}^{\textsc{gut}} &=& \frac{1}{\Lambda^2}
({\cal D}^\alpha H)^{\dagger\,}\overline{D}_R^{} \Big( Y_d^\dagger\Delta_{q1}^{} +
Y_e^*\Delta_{q2}^{} + \Delta_{\ell3}^{\textsc t}Y_d^\dagger + \Delta_{\ell4}^{\textsc t}Y_e^* +
\Delta_{\ell3}^{{\scriptscriptstyle\prime}\textsc t}Y_d^\dagger
\Delta_{q1}^{\scriptscriptstyle\prime} +
\Delta_{\ell4}^{{\scriptscriptstyle\prime}\textsc t}Y_e^*\Delta_{q2}^{\scriptscriptstyle\prime}
\Big) {\cal D}_\alpha^{}Q_L^{}
\nonumber \\ && \! +\;
\frac{1}{\Lambda^2}
({\cal D}^\alpha H)^{\dagger\,} \overline{E}_R^{} \Big( Y_e^\dagger\Delta_{\ell1}^{} +
Y_d^*\Delta_{\ell2}^{} + \Delta_{q3}^{\textsc t}Y_d^* + \Delta_{q4}^{\textsc t}Y_e^\dagger +
\Delta_{q3}^{{\scriptscriptstyle\prime}\textsc t}Y_d^*\Delta_{\ell2}^{\scriptscriptstyle\prime} +
\Delta_{q4}^{{\scriptscriptstyle\prime}\textsc t}Y_e^\dagger
\Delta_{\ell1}^{\scriptscriptstyle\prime}
\Big) {\cal D}_\alpha^{}L_L^{}
\nonumber \\ && \! +\; {\rm H.c.} \,,
\end{eqnarray}
where $\Delta_{qn}^{\scriptscriptstyle(\prime)}$ and
$\Delta_{\ell n}^{\scriptscriptstyle(\prime)}$ are the same in form as $\Delta_q$ and
$\Delta_\ell$, respectively, but have their own coefficients
$\zeta_r^{\scriptscriptstyle(\prime)}$ and $\xi_r^{\scriptscriptstyle(\prime)\,}$ $(r=1,2,4)$.
We notice that, while the $\Delta_{q1}$ and $\Delta_{\ell1}$ terms in
${\cal L}_{\textsc{mfv}}^{\textsc{gut}}$ already occur in the non-GUT case, Eq.\,(\ref{Lmfv}),
the others are new here.
In general, the different quark and lepton operators in Eq.\,(\ref{Lgutmfv0}) may be
unrelated to each other, depending on the specifics of the underlying model, and so it is
possible that only one or some of the terms in ${\cal L}_{\textsc{mfv}}^{\textsc{gut}}$
dominate the nonstandard contribution to \,$h\to f\bar f'$.\,
Therefore, we will consider different possible scenarios below.
As in the non-GUT framework of the last section, we will evaluate the contributions of
${\cal L}_{\textsc{mfv}}^{\textsc{gut}}$ to Higgs decay model-independently and not deal with
the constraints on the GUT-MFV counterparts of the operators in Eqs.\,\,(\ref{others-q})
and\,\,(\ref{others-l}), as the potential links among the $\Delta$s belonging to
these various operators again depend on model details.

Working in the mass eigenstate basis, we derive from Eq.\,(\ref{Lgutmfv0})
\begin{eqnarray} \label{Lgutmfv}
{\cal L}_{\textsc{mfv}}^{\textsc{gut}} &\supset&
\frac{\partial^\alpha h}{\sqrt2\,\Lambda^2}\,\overline{D}_R^{} \!\begin{array}[t]{l}
\big( Y_d^{}\Delta_{q1}^{} + \textsf{G}^\dagger Y_e^{}\textsf{C}_{\,}\Delta_{q2}^{}
+ \textsf{G}^\dagger\Delta_{\ell3}^{\textsc t\;}\textsf{G}_{\,} Y_d^{}
+ \textsf{G}^\dagger\Delta_{\ell4}^{\textsc t\;} Y_e^{}\textsf{C}
\medskip \\
\;+\; \textsf{G}^\dagger\Delta_{\ell3}^{{\scriptscriptstyle\prime}\textsc t\;}\textsf{G}_{\,}Y_d^{}
\Delta_{q1}^{\scriptscriptstyle\prime} +
\textsf{G}^\dagger\Delta_{\ell4}^{{\scriptscriptstyle\prime}\textsc t}Y_e^{}\textsf{C}_{\,}
\Delta_{q2}^{\scriptscriptstyle\prime} \big) \partial_\alpha^{}D_L^{}
\end{array}
\nonumber \\ && \! +\;
\frac{\partial^\alpha h}{\sqrt2\,\Lambda^2}\, \overline{E}_R^{} \!\begin{array}[t]{l}
\big( Y_e^{}\Delta_{\ell1}^{} + \textsf{C}^* Y_d^{\;}\textsf{G}^{\textsc t}\Delta_{\ell2}^{}
+ \textsf{C}^*\Delta_{q3}^{\textsc t} Y_d^{\;}\textsf{G}^{\textsc t}
+ \textsf{C}^*\Delta_{q4}^{\textsc t}\textsf{C}^{\textsc t}Y_e^{}
\medskip \\
\;+\; \textsf{C}^*\Delta_{q3}^{{\scriptscriptstyle\prime}\textsc t}Y_d^{\;}\textsf{G}^{\textsc t}
\Delta_{\ell2}^{\scriptscriptstyle\prime} +
\textsf{C}^*\Delta_{q4}^{{\scriptscriptstyle\prime}\textsc t}\textsf{C}^{\textsc t}Y_e^{}
\Delta_{\ell1}^{\scriptscriptstyle\prime} \big) \partial_\alpha^{} E_L^{}
\end{array}
\nonumber \\ && \! +\; {\rm H.c.} \,,
\end{eqnarray}
where now the column matrices $D_{L,R}$ and $E_{L,R}$ contain mass eigenstates, $Y_{d,e}$ are
diagonal and real as in Eq.\,(\ref{YdYe}), the formulas for ${\sf A}_{q,\ell}$ in
$\Delta_{qn,\ell n}^{\scriptscriptstyle(\prime)}$, respectively, are
those in Eq.\,(\ref{BqBl}), and
\begin{eqnarray} \label{CG}
\textsf{C} \,=\, {\cal V}_{e_R}^{\textsc t}{\cal V}_{d_L}^{} \,, ~~~~ ~~~
\textsf{G} \,=\, {\cal V}_{e_L}^{\textsc t}{\cal V}_{d_R}^{} \,,
\end{eqnarray}
with ${\cal V}_{d_L,d_R}$ and ${\cal V}_{e_L,e_R}$ being the unitary matrices in
the biunitary transformations that diagonalize $Y_d$ and $Y_e$, respectively.
Since the elements of ${\cal V}_{d_L,d_R}$ and ${\cal V}_{e_L,e_R}$ are unknown,
so are those of $\textsf{C}$ and $\textsf{G}$.
Nevertheless, it has been pointed out in Ref.\,\,\cite{Grinstein:2006cg} that the two
matrices have hierarchical textures.
As indicated in Appendix\,\ref{app}, this implies that the limit
\,$\textsf{C}=\textsf{G}=\openone$\, is one possibility that may be entertained for
order-of-magnitude considerations~\cite{Grinstein:2006cg,Filipuzzi:2009xr}.
It corresponds to neglecting the subdominant $\lambda_5'$ contributions in Eq.\,(\ref{Ygut}).
Due to the lack of additional information about $\textsf{C}$ and $\textsf{G}$,
in what follows we concentrate on this special scenario for simplicity, in which case
the Yukawa couplings from Eq.\,(\ref{Lgutmfv}) are
\begin{eqnarray}
{\cal Y}_{D_kD_l}^{} &=& {\cal Y}_{D_kD_{l}}^{\textsc{sm}}
\,-\, \frac{m_h^2}{2\Lambda^2v} \Big[ \big(\Delta_{q1}\big)_{kl\,}m_{D_l}^{} +
\big(\Delta_{q2}\big)_{kl\,} m_{E_l}^{} + m_{D_k}^{} \big(\Delta_{\ell3}\big)_{lk}
+ m_{E_k}^{} \big(\Delta_{\ell4} \big)_{lk} \Big]
\nonumber \\ && \!\! -\;
\frac{m_h^2}{2\Lambda^2v} \big( \Delta_{q1}^{\scriptscriptstyle\prime}\hat M_d^{~}
\Delta_{\ell3}^{{\scriptscriptstyle\prime}\textsc t} +
\Delta_{q2}^{\scriptscriptstyle\prime}\hat M_e^{~}
\Delta_{\ell4}^{{\scriptscriptstyle\prime}\textsc t} \big)_{kl} \,, \vphantom{\int_{\int_|^|}}
\nonumber \\
{\cal Y}_{E_kE_l}^{} &=& \delta_{kl}^{}\,{\cal Y}_{E_kE_k}^{\textsc{sm}}
\,-\, \frac{m_h^2}{2\Lambda^2v} \Big[ \big(\Delta_{\ell1}\big)_{kl\,} m_{E_l}^{}
+ \big(\Delta_{\ell2}\big)_{kl\,} m_{D_l}^{} + m_{D_k}^{} \big(\Delta_{q3}\big)_{lk\,}
+ m_{E_k}^{} \big(\Delta_{q4}\big)_{lk} \Big]
\nonumber \\ && \!\! -\;
\frac{m_h^2}{2\Lambda^2v} \big( \Delta_{\ell2}^{\scriptscriptstyle\prime}\hat M_d^{}
\Delta_{q3}^{{\scriptscriptstyle\prime}\textsc t} +
\Delta_{\ell1}^{\scriptscriptstyle\prime}\hat M_e^{}
\Delta_{q4}^{{\scriptscriptstyle\prime}\textsc t} \big)_{kl} \,, \label{YEkElgut}
\end{eqnarray}
where \,$\hat M_d^{}=Y_d^{~}v/\sqrt2={\rm diag}(m_d,m_s,m_b)$\, and
\,$\hat M_e^{}=Y_e^{~}v/\sqrt2={\rm diag}(m_e,m_\mu,m_\tau)$.\,

To gain some insight into the potential impact of the new terms on these Yukawas,
we can explore several different simple scenarios in which only one or more of the $\Delta$s
are nonvanishing.
If $\Delta_{q1}$ and $\Delta_{\ell1}$ are the only ones present and independent of each
other, their effects are the same as those of $\Delta_q$ and $\Delta_\ell$, respectively,
investigated in the previous section and Ref.\,\,\cite{He:2015rqa}.
In the rest of this section, we look at other possible cases.

In the first one, we assume that $\Delta_{\ell2}$ is the only new source in Eq.\,(\ref{YEkElgut}).
In view of the rough similarity between the $\Delta_{\ell1}$ and $\Delta_{\ell2}$ portions of
${\cal Y}_{E_kE_l}$, due to \,$m_\mu^{}/m_s^{}\sim m_b^{}/m_\tau^{}\sim 2$\, at
the renormalization scale \,$\mu\sim m_h^{}/2$,\, we can infer that the situation in this case is
not much different from its $\Delta_\ell$ counterpart addressed briefly in the last section and
treated more extensively in Ref.\,\,\cite{He:2015rqa}.
In other words, for the $\Delta_{\ell2}$ term alone to achieve \,${\cal B}(h\to\mu\tau)\sim1$\%\,
and meet the other requirements described earlier simultaneously, the $O$ matrix occurring
in $\textsf{A}_\ell$, as defined in\,\,Eq.\,(\ref{BqBl}), must be complex in order to
provide the extra free parameters needed to raise $|{\cal Y}_{\mu\tau}|$ and reduce
$|{\cal Y}_{e\mu}|$ sufficiently.
If $\Delta_{\ell1}$ is also nonvanishing and equals $\Delta_{\ell2}$, the picture is
qualitatively unchanged.
We have verified all this numerically.

Still another possibility with $\Delta_{\ell n}$ is that all the $\Delta_{q n}$
are absent and that ${\cal Y}_{D_kD_l}$ and ${\cal Y}_{E_kE_l}$ each have at least
one $\Delta_{\ell n}$.
In this case, if, say, only $\Delta_{\ell1,\ell3}$ are present and
\,$\Delta_{\ell1}=\Delta_{\ell3}$,\, we find that it is not possible to reach
the desired \,$|{\cal Y}_{\mu\tau}|>0.002$\, and satisfy the constraints in
the quark sector at the same time.
The situation is not improved by keeping all the $\Delta_{\ell n}$, while still taking
them to be equal.
However, if the $\Delta_{\ell n}$ contributions to ${\cal Y}_{D_kD_l}$ are weakened
by an overall factor of 2 or more, at least part of the requisite range of
$|{\cal Y}_{\mu\tau}|$ can be attained.

\begin{figure}[b] \smallskip
\includegraphics[width=7cm]{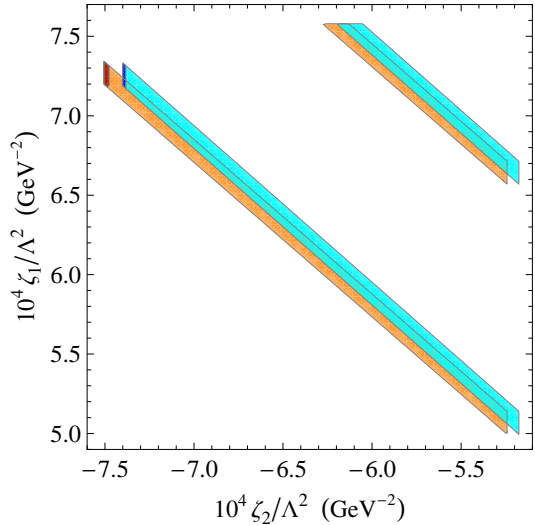} \vspace{-1ex}
\caption{Regions of $\zeta_1^{}/\Lambda^2$ and $\zeta_2^{}/\Lambda^2$ for \,$\zeta_4^{}=0$\,
(cyan and dark blue) which satisfy the experimental constraints in
Eqs.\,(\ref{|Yll'|limit})-(\ref{|Yll|limit}) if the $\Delta_{q3}$ term is the only
new-physics contribution in Eq.\,(\ref{YEkElgut}).
For the orange and dark red regions, the roles of $\zeta_2^{}$ and $\zeta_4^{}$ are interchanged.
The dark (blue and red) patches correspond to \,$|{\cal Y}_{\tau\mu}|\simeq0.0029$\,
and hence \,${\cal B}(h\to\mu\tau)\simeq1$\%.\label{z1z2-ll}} \vspace{-2ex}
\end{figure}

An interesting case is where $\Delta_{q3}$ is nonvanishing and all of the other
$\Delta$s in Eq.\,(\ref{YEkElgut}) are absent.
This implies that the flavor changes depend entirely on the known CKM parameters and quark masses.
Furthermore, \,$|{\cal Y}_{\mu e,\tau e,\tau\mu}|\gg|{\cal Y}_{e\mu,e\tau,\mu\tau}|$,\,
respectively, as can be deduced from Eq.\,(\ref{YEkElgut}).
It turns out that the leptonic restrictions in Eqs.\,(\ref{|Yll'|limit})-(\ref{|Yll|limit})
can be satisfied together with only the $\zeta_1^{}$ and $\zeta_2^{}$, or $\zeta_4^{}$, terms
in $\Delta_{q3}$ being present.
We also find that the largest $|{\cal Y}_{\tau\mu}|$ that can be attained is
\,{\small$\sim$\,}0.0029.\,
We illustrate this in Fig.\,\ref{z1z2-ll}, where the cyan and dark blue (orange and dark red)
areas correspond to only $\zeta_{1,2}^{}$  $\big(\zeta_{1,4}^{}\big)$ being nonzero.
The widths of the two (colored) bands in this graph are controlled by the ${\cal Y}_{\tau\tau}$
constraint, whereas the vertical and horizontal ranges are restrained by Eq.\,(\ref{|Ymt|limit})
as well as the ${\cal Y}_{\mu\mu}$ constraint and Eq.\,(\ref{m2eglimit}).
To show some more details of this case, we collect in Table\,\,\ref{yukawas} a few sample
values of the Yukawa couplings in the allowed parameter space.
Evidently, the predictions on ${\cal Y}_{\mu\mu,\tau\tau}$ can deviate markedly from their
SM values and, therefore, will likely be confronted with more precise measurements of
\,$h\to\mu^+\mu^-,\tau^+\tau^-$\, in the near future.
As expected, the flavor-violating couplings obey the magnitude ratio
\,$|{\cal Y}_{\mu e}|:|{\cal Y}_{\tau e}|:|{\cal Y}_{\tau\mu}|\simeq
|({\sf A}_q)_{12}^{}|m_s^{}:|({\sf A}_q)_{13}^{}|m_b^{}:
|({\sf A}_q)_{23}^{}|m_b^{}\simeq0.00017:0.21:1$,\,
compatible with Eq.\,(\ref{Yqq'ratio}).
Also listed in the table are the branching fractions of the decay \,$\mu\to e\gamma$\, and
\,$\mu\to e$\, conversion in aluminum nuclei, computed with the formulas
collected in Ref.\,\,\cite{He:2015rqa} under the assumption that these transitions are
induced by the Yukawas alone.
The \,$\mu\to e\gamma$\, numbers are below the current experimental bound
\,${\cal B}(\mu\to e\gamma)<4.2\times10^{-13}$ \cite{meg}, but not by very much.
Hence they will probably be checked by the planned MEG\,II experiment with sensitivity
anticipated to reach a\,\,few times $10^{-14}$ after 3 years of data taking~\cite{CeiA:2014wea}.
Complementarily, the ${\cal B}(\mu_{\,\!}{\rm Al}\to e_{\,\!}{\rm Al})$ results can be probed
by the upcoming Mu2E and COMET searches, which utilize aluminum as the target material and
are expected to have sensitivity levels under $10^{-16}$ after several years of
running~\cite{CeiA:2014wea}.

\begin{table}[t]
\begin{tabular}{|cccccc|cc|} \hline
\,$\displaystyle\frac{{\cal Y}_{ee}^{}}{{\cal Y}_{ee_{\vphantom{\int}}}^{\textsc{sm}}}$\,
& ~$\displaystyle\frac{{\cal Y}_{\mu\mu}^{}}{{\cal Y}_{\mu\mu}^{\textsc{sm}}}$\,
& ~$\displaystyle\frac{{\cal Y}_{\tau\tau}^{}}{{\cal Y}_{\tau\tau}^{\textsc{sm}}}$~
& $\displaystyle\frac{{\cal Y}_{\mu e}^{}}{10\raisebox{0.3pt}{$^{-7}$}}$
& $\displaystyle\frac{{\cal Y}_{\tau e}^{}}{10\raisebox{0.3pt}{$^{-4}$}}$
& $\displaystyle\frac{{\cal Y}_{\tau\mu}^{\vphantom{\int}}}{10\raisebox{0.3pt}{$^{-3}$}}$
& \,${\cal B}(\mu\to e\gamma)$\, & \,${\cal B}(\mu_{\,\!}{\rm Al}\to e_{\,\!}{\rm Al})$\,
\\ \hline\hline
$-$31 & $-$2.1 & 0.95 & ~$-4.3-1.9i$~ & ~$5.5+2.3i$~ & ~$-2.8+0.05i$~ & ~$4.0\times10^{-13}$~ &
$2.0\times10^{-15}\vphantom{|_|^{\int}}$
\\
$-$28 & $-$1.8 & 1.1  &  $-4.0-1.7i$  &  $5.1+2.1i$  &  $-2.6+0.05i$  &  $3.1\times10^{-13}$  &
$1.6\times10^{-15}\vphantom{|_|^|}$
\\
$-$24 & $-$1.4 & 1.0  &  $-3.4-1.5i$  &  $4.3+1.8i$  &  $-2.2+0.04i$  &  $1.7\times10^{-13}$  &
$9.5\times10^{-16}\vphantom{|_o^|}$
\\
\hline
\end{tabular}
\caption{Higgs-lepton Yukawa couplings if the $\Delta_{q3}$ term with \,$\zeta_4^{}=0$\, is
the only new-physics contribution in Eq.\,(\ref{YEkElgut}), and the resulting branching fractions
of the \,$\mu\to e\gamma$\, decay and \,$\mu\to e$\, conversion in aluminum nuclei.\label{yukawas}}
\vspace{-1ex} \end{table}

In contrast to the preceding paragraph, if $\Delta_{q4}$ instead of $\Delta_{q3}$ is
nonvanishing and the other $\Delta$s remain absent, the desired size of $|{\cal Y}_{\tau\mu}|$
becomes unattainable, as it can be at most \,{\footnotesize$\sim$\,}0.001,\,
even with $\zeta_{1,2,4}^{}$ being nonzero.
If both $\Delta_{q3,q4}$ are the only ones present and they are identical, we find
\,$|{\cal Y}_{\tau\mu}|\sim0.0017$\, to be the biggest achievable, somewhat below the lower
limit in Eq.\,(\ref{|Ymt|limit}).

If instead $\Delta_{q1}$ and $\Delta_{q3}$ are the only ones nonvanishing and
\,$\Delta_{q1}=\Delta_{q3}$,\, the quark sector constraints in
Eqs.\,(\ref{|Yqq'|limits})-(\ref{|Ybb|limit}) do not permit
$|{\cal Y}_{\tau\mu}|$ to exceed 0.00072, which is almost 3 times less than
the required minimum in Eq.\,(\ref{|Ymt|limit}).
This implies that, alternatively, if the $\Delta_{q1}$ contribution to ${\cal Y}_{D_kD_l}$ is
decreased by an overall factor of 3 or more, at least part of the desired
$|{\cal Y}_{\tau\mu}|$ range can be reached and the other restrictions fulfilled.

Lastly, we look at the $\Delta_{\ell2}^{\scriptscriptstyle\prime}\hat M_d^{}
\Delta_{q3}^{{\scriptscriptstyle\prime}\textsc t}$ and
$\Delta_{\ell1}^{\scriptscriptstyle\prime}\hat M_e^{}
\Delta_{q4}^{{\scriptscriptstyle\prime}\textsc t}$ parts in ${\cal Y}_{E_kE_l}$.
With
\,$\Delta_{q3}^{\scriptscriptstyle\prime}=\zeta_1^{\scriptscriptstyle\prime}\openone +
\zeta_2^{\scriptscriptstyle\prime}{\sf A}_q^{} +
\zeta_4^{\scriptscriptstyle\prime}{\sf A}_q^2$\, and
\,$\Delta_{\ell2}^{\scriptscriptstyle\prime}=\xi_1^{\scriptscriptstyle\prime}\openone +
\xi_2^{\scriptscriptstyle\prime}{\sf A}_\ell^{} +
\xi_4^{\scriptscriptstyle\prime}{\sf A}_\ell^2$,\,
using in particular ${\sf A}_q$ from Eq.\,(\ref{Aq}) and ${\sf A}_\ell$
from Eq.\,(\ref{AlNH}), we see that
$\Delta_{\ell2}^{\scriptscriptstyle\prime}\hat M_d^{}
\Delta_{q3}^{{\scriptscriptstyle\prime}\textsc t}$ has two more free parameters,
$\zeta_{2,4}^{\scriptscriptstyle\prime}$
$\big(\xi_{^{\scriptstyle2,4}}^{\scriptscriptstyle\prime}\big)$, compared
to $\Delta_{\ell2}\hat M_d$ $\big(\hat M_d\Delta_{^{\scriptstyle q3}}^{\textsc t}\big)$.
It turns out, however, that the presence of additional parameters does not necessarily
translate into more freedom for
the $\Delta_{\ell2}^{\scriptscriptstyle\prime}\hat M_d^{}
\Delta_{q3}^{{\scriptscriptstyle\prime}\textsc t}$ contributions due to the following reason.
With $\hat M_d$ being sandwiched between $\Delta_{\ell2}^{\scriptscriptstyle\prime}$ and
$\Delta_{q3}^{{\scriptscriptstyle\prime}\textsc t}$, in general ${\cal Y}_{ff'}$ for
\,$f\neq f'$\, can be comparable in size to ${\cal Y}_{f'f}$ because they both have terms
linear in $m_b^{}$, as do ${\cal Y}_{ee,\mu\mu}$, which is unlike the situation of
the ${\cal Y}_{E_kE_l}$ parts containing only one $\Delta$.
We find that, once the two extra free parameters are fixed to suppress the $m_b^{}$ effects on
\,$\mu\to e\gamma$\, as well as \,$h\to\mu^+\mu^-$,\, the predictions for the various
${\cal Y}_{E_kE_l}$ are not very different qualitatively from those in the $\Delta_{\ell2}$
$(\Delta_{q3})$ case examined earlier.
Similarly, the implications of the contributions of
$\Delta_{\ell1}^{\scriptscriptstyle\prime}\hat M_e^{}
\Delta_{q4}^{{\scriptscriptstyle\prime}\textsc t}$ do not differ much from those of
$\Delta_{\ell1}\hat M_e$ or $\hat M_e\Delta_{q4}^{\textsc t}$ also discussed earlier.

The above simple scenarios have specific predictions for the flavor-conserving and -violating
Yukawa couplings and hence are all potentially testable in upcoming measurements of
\,$h\to f\bar f'$\, and searches for flavor-violating charged-lepton transitions such as
\,$\mu\to e\gamma$.\,
If the predictions disagree with the collected data, more complicated cases could be proposed
in order to probe further the GUT MFV framework that we have investigated.

\section{Conclusions\label{conclusion}}

We have explored the flavor-changing decays of the Higgs boson into down-type fermions
in the MFV framework based on the SM extended with the addition of right-handed neutrinos plus
effective dimension-six operators and in its SU(5) GUT counterpart.
As a consequence of the MFV hypothesis being applied in the latter framework, we are able to
entertain the possibility that the recent tentative indication of \,$h\to\mu\tau$\, in
the LHC data has some connection with potential new physics in the quark sector.
Here the link is realized specifically by leptonic (quark) bilinears involving quark (leptonic)
Yukawa combinations that control the leptonic (quark) flavor changes.
We discuss different simple scenarios in this context and how they are subject to various
experimental requirements.
In one particular case, the leptonic Higgs couplings are determined mainly by the known CKM
parameters and quark masses, and interestingly their current values allow the couplings to yield
\,${\cal B}(h\to\mu\tau)\sim1$\%\, without being in conflict with other constraints.
Forthcoming measurements of the Higgs fermionic decays and searches for flavor-violating
charged-lepton decays will expectedly provide extra significant tests on the GUT MFV scenarios
studied here.

\acknowledgments

The work of J.T. was supported in part by the MOE Academic Excellence Program
(Grant No.\,102R891505) of Taiwan.
He would like to thank S.B. and Pyungwon Ko for generous hospitality at the Korea
Institute for Advanced Study during the course of this research.
This work is supported in part by National Research Foundation of Korea (NRF) Research Grant
NRF-2015R1A2A1A05001869 (S.B.).

\appendix

\section{\boldmath$\sf C$ and $\sf G$ matrices\label{app}}

The unitary matrices $\sf C$ and $\sf G$ defined in Eq.\,(\ref{CG}) have unknown elements,
but are expected to be hierarchical in structure~\cite{Grinstein:2006cg}.
Expressing each of them as an expansion in the Wolfenstein parameter \,$\lambda\simeq0.23$,\,
we have
\begin{eqnarray} \label{cg}
\textsf{C}\,=\,& \left(\!\begin{array}{ccc} {\tt C}_{11}^{} & {\tt C}_{12}^{} &
\lambda^2 ({\tt C}_{11}^{}\textsc{c}_1^{}+{\tt C}_{12}^{}\textsc{c}_2^{}) \vspace{3pt} \\
{\tt C}_{21}^{} & {\tt C}_{22}^{} &
\lambda^2 ({\tt C}_{21}^{}\textsc{c}_1^{}+{\tt C}_{22}^{}\textsc{c}_2^{}) \vspace{3pt} \\
-\lambda^2 \textsc{c}_1^* & \mbox{$-\lambda^2\textsc{c}_2^*$} & 1 \end{array}\!\right) ,
\nonumber \\
\textsf{G} \,=\,& \left(\!\begin{array}{ccc} {\tt G}_{11}^{} & {\tt G}_{12}^{} &
\lambda^2 ({\tt G}_{11}^{}\textsc{g}_1^{}+{\tt G}_{12}^{}\textsc{g}_2^{}) \vspace{3pt} \\
{\tt G}_{21}^{} & {\tt G}_{22}^{} &
\lambda^2 ({\tt G}_{21}^{}\textsc{g}_1^{}+{\tt G}_{22}^{}\textsc{g}_2^{}) \vspace{3pt} \\
-\lambda^2 \textsc{g}_1^* & \mbox{$-\lambda^2\textsc{g}_2^*$} & 1 \end{array}\!\right) ~~
\end{eqnarray}
up to order $\lambda^3$, where \,${\tt C}_{ac}$, $\textsc{c}_a^{}$, ${\tt G}_{ac}$, and
$\textsc{g}_a^{}$\, are parameters
with magnitudes below 1 and we have used the approximation \,$y_\mu^{}/y_\tau^{}\sim\lambda^2$.\,
For discussion purposes, it suffices to look at only two of the flavor-violating matrix combinations
occurring in Eq.\,(\ref{Lgutmfv}), namely $\textsf{G}^\dagger Y_e^{}\textsf{C}_{\,}{\sf A}_q^{}$
and $\textsf{C}^*{\sf A}_q^{\textsc t} Y_d^{}\textsf{G}^{\textsc t}$ which are parts of
$\textsf{G}^\dagger Y_e^{}\textsf{C}_{\,}\Delta_{q2}^{}$
and $\textsf{C}^*\Delta_{q3}^{\textsc t} Y_d^{}\textsf{G}^{\textsc t}$, respectively.
Expanding their matrix elements in\,\,$\lambda$, we express these combinations as
\begin{align}
\textsf{G}^\dagger Y_e^{}\textsf{C}_{\,}{\sf A}_q^{} \,=\,& \left(\begin{array}{ccc}
{\cal O}\big(\lambda^5\big)\,y_\tau^{} & ~ {\cal O}\big(\lambda^4\big)\,y_\tau^{} ~ &
\lambda^2 [{\tt C}_{21}^{}\textsc{c}_1^{}+{\tt C}_{22}^{}(\textsc{c}_2^{}-A)]{\tt G}_{21\,}^*y_\mu^{}
- \lambda^2\textsc{g}_{1\,}^{}y_\tau^{} \vspace{3pt} \\
{\cal O}\big(\lambda^5\big)\,y_\tau^{} & {\cal O}\big(\lambda^4\big)\,y_\tau^{} &
\lambda^2 [{\tt C}_{21}^{}\textsc{c}_1^{}+{\tt C}_{22}^{}(\textsc{c}_2^{}-A)]{\tt G}_{22\,}^*y_\mu^{}
- \lambda^2\textsc{g}_{2\,}^{}y_\tau^{} \vspace{3pt} \\
\lambda^3 A_{\,}(1-\rho-i\eta)y_\tau^{} & -\lambda^2 A_{\,}y_\tau^{} & y_\tau^{} \end{array}\right) ,
\nonumber \\
\textsf{C}^*{\sf A}_q^{\textsc t} Y_d^{}\textsf{G}^{\textsc t} \,=\,& \left(\begin{array}{ccc}
{\cal O}\big(\lambda^4\big) & {\cal O}\big(\lambda^4\big) &
\lambda^2 [{\tt C}_{11}^*\textsc{c}_1^*+{\tt C}_{12}^*(\textsc{c}_2^*-A)]
\vspace{3pt} \\
{\cal O}\big(\lambda^4\big) & {\cal O}\big(\lambda^4\big) &
\lambda^2 [{\tt C}_{21}^*\textsc{c}_1^*+{\tt C}_{22}^*(\textsc{c}_2^*-A)]
\vspace{3pt} \\
\lambda^2 ({\tt G}_{11}^{}\textsc{g}_1^{}+{\tt G}_{12}^{}\textsc{g}_2^{}) & ~\lambda^2
({\tt G}_{21}^{}\textsc{g}_1^{}+{\tt G}_{22}^{}\textsc{g}_2^{}) ~ & 1
\end{array}\right) \! y_b^{} \,, \label{CAqYdG}
\end{align}
where we have kept only $y_{\tau,b}^{}$ terms to the leading nonzero order in $\lambda$ and
$y_\mu^{}$ terms to order $\lambda^2$, made use of \,$y_c^2/y_t^2\sim2\lambda^8$\, and
\,$y_s^{}/y_b^{}\sim2\lambda^3$,\, and set \,$y_t^{}=1$.\,
Being unknown, one or more of ${\tt C}_{ac}$ and $\textsc{c}_a^{}$ may be small or vanishing,
although the unitarity of $\sf C$ implies
\begin{eqnarray}
|{\tt C}_{11}|^2+|{\tt C}_{12}|^2 \,=\, 1 \,, ~~~~~ |{\tt C}_{11}| \,=\, |{\tt C}_{22}| \,, ~~~~
|{\tt C}_{12}| \,=\, |{\tt C}_{21}| \,, ~~~~
{\tt C}_{11}^{}{\tt C}_{12}^* \,=\, -{\tt C}_{21}^{}{\tt C}_{22}^* \,,
\end{eqnarray}
valid to order $\lambda^2$.  The same can be said of the elements of $\sf G$.
It follows that we may choose \,$\textsf{C}=\textsf{G}=\openone$\, as a possible limit for these
matrices~\cite{Grinstein:2006cg,Filipuzzi:2009xr},
in which case Eq.\,(\ref{CAqYdG}) becomes
\begin{align}
Y_e^{}{\sf A}_q^{} \,=\,& \left(\begin{array}{ccc}
0 & 0 & 0 \vspace{3pt} \\ 0 & 0 & -\lambda^2 A_{\,}y_\mu^{} \vspace{3pt} \\
\lambda^3 A_{\,}(1-\rho-i\eta)_{\,}y_\tau^{} & \mbox{~ $-\lambda^2 A_{\,}y_\tau^{}$ ~} &
y_\tau^{} \end{array}\right) +\, {\cal O}\big(\lambda^5\big) \,,
\nonumber \\
{\sf A}_q^{\textsc t} Y_d^{} \,=\,& \left(\begin{array}{ccc} 0 & ~~ 0 ~~ &
\lambda^3 A_{\,}(1-\rho-i\eta)_{\,}y_b^{} \vspace{3pt} \\
0 & 0 & -\lambda^2 A_{\,}y_b^{} \vspace{3pt} \\ 0 & 0 & y_b^{} \end{array}\right)
+\, {\cal O}\big(\lambda^5\big) \,.
\end{align}
Taking this limit corresponds to neglecting the nonleading $\epsilon\lambda_5'$ terms in
Eq.\,(\ref{Ygut}) which break the \,$Y_d^{}=Y_e^{\textsc t}$ relation
($\textsf{C},\textsf{G}\to\openone$\, if \,$\epsilon\to0$) and simplifies the treatment of
quantities that depend on $\textsf{C}$ and $\textsf{G}$.
However, since not much is known about their elements, their presence precludes a\,\,precise
evaluation of such quantities~\cite{Grinstein:2006cg}.
The implication is that the results of our GUT MFV calculations involving the Yukawas
with \,$\textsf{C}=\textsf{G}=\openone$\, from Eq.\,(\ref{YEkElgut})
should be understood as only order-of-magnitude estimates.

\end{document}